# TECHNIQUES FOR FEATURE EXTRACTION IN SPEECH RECOGNITION SYSTEM : A COMPARATIVE STUDY

*Urmila Shrawankar*
Research Student, (Computer Science & Engg.), SGB Amravati University
urmilas@rediffmail.com

*Dr. Vilas Thakare*
Professor & Head, PG Dept. of Computer Science, SGB Amravati University, Amravati

**ABSTRACT**
*The time domain waveform of a speech signal carries all of the auditory information. From the phonological point of view, very little can be said on the basis of the waveform itself. However, past research in mathematics, acoustics, and speech technology have provided many methods for converting data, that can be considered as information if interpreted correctly. In order to find some statistically relevant information from incoming data, it is important to have mechanisms for reducing the information of each segment in the audio signal into a relatively small number of parameters, or features. These features should describe each segment in such a characteristic way that other similar segments can be grouped together by comparing their features. There are enormous interesting and exceptional ways to describe the speech signal in terms of parameters. Though, they all have their strengths and weaknesses, we have presented some of the most used methods with their importance.*

## I. INTRODUCTION

Speech is one of the ancient ways to express ourselves. Today these speech signals are also used in biometric recognition technologies and communicating with machine.

These speech signals are slowly timed varying signals (quasi-stationary). When examined over a sufficiently short period of time (5-100 msec), its characteristics are fairly stationary. But, if for a period of time the signal characteristics changes, it reflects to the different speech sounds being spoken. The information in speech signal is actually represented by short term amplitude spectrum of the speech wave form. This allows us to extract features based on the short term amplitude spectrum from speech (phonemes).

The fundamental difficulty of speech recognition is that the speech signal is highly variable due to different speakers, nt speaking rates, contents and acoustic conditions.

The feature analysis component of an ASR system plays a crucial role in the overall performance of the system. Many feature extraction techniques are available, these include

- Linear predictive analysis (LPC)
- Linear predictive cepstral coefficients (LPCC),
- perceptual linear predictive coefficients (PLP)
- Mel-frequency cepstral coefficients (MFCC)
- Power spectral analysis (FFT)
- Mel scale cepstral analysis (MEL)
- Relative spectra filtering of log domain coefficients (RASTA)
- First order derivative (DELTA)

Etc.

## II. DIGITAL SIGNAL PROCESSING (DSP) TECHNIQUES

The signal processing front-end, which converts the speech waveform to some of type of parametric representation. This parametric representation is then used for further analysis and processing.

Digital signal processing (DSP) techniques is the core of speech recognition system. DSP methods are used in speech analysis, synthesis, coding, recognition, and enhancement, as well as voice modification, speaker recognition, and language identification.

## III. FEATURE EXTRACTION

Theoretically, it should be possible to recognize speech directly from the digitized waveform. However, because of the large variability of the speech signal, it is better to perform some feature extraction that would reduce that variability. Particularly, eliminating various source of information, such as whether the sound is voiced or unvoiced and, if voiced, it eliminates the effect of the periodicity or pitch, amplitude of excitation signal and fundamental frequency etc.

The reason for computing the short-term spectrum is that the cochlea of the human ear performs a quasi-frequency analysis. The analysis in the cochlea takes place on a nonlinear frequency scale (known as the Bark scale or the mel scale). This scale is approximately linear up to about 1000 Hz and is approximately logarithmic thereafter. So, in the feature extraction, it is very common to perform a frequency warping of the frequency axis after the spectral computation.

This section is a summary of feature extraction techniques that are in use today, or that may be useful in the future, especially in the speech recognition area. Many of these techniques are also useful in other areas of speech processing.

## IV. LINEAR PREDICTIVE CODING (LPC)

LPC is one of the most powerful speech analysis techniques and is a useful method for encoding quality speech at a low bit rate. The basic idea behind linear predictive analysis is that a specific speech sample at the current time can be approximated as a linear combination of past speech samples.

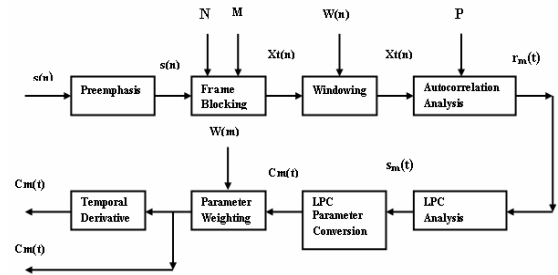

Fig : LPC : The LPC processor

**Methodology**

LP is a model based on human speech production. It utilizes a conventional source-filter model, in which the glottal, vocal tract, and lip radiation transfer functions are integrated into one all-pole filter that simulates acoustics of the vocal tract.

The principle behind the use of LPC is to minimize the sum of the squared differences between the original speech signal and the estimated speech signal over a finite duration. This could be used to give a unique set of predictor coefficients. These predictor coefficients are estimated every frame, which is normally 20 ms long. The predictor coefficients are represented by $a_k$. Another important parameter is the gain (G). The transfer function of the time varying digital filter is given by

$$H(z) = G/(1-\Sigma a_k z^{-k})$$

Where k=1 to p, which will be 10 for the LPC-10 algorithm and 18 for the improved algorithm that is utilized. Levinsion-Durbin recursion will be utilized to compute the required parameters for the auto-correlation method (Deller et al., 2000).

The LPC analysis of each frame also involves the decision-making process of voiced or unvoiced. A pitch-detecting algorithm is employed to determine to correct pitch period / frequency. It is important to re-emphasis that the pitch, gain and coefficient parameters will be varying with time from one frame to another.

In reality the actual predictor coefficients are never used in recognition, since they typical show high variance. The predictor coefficient is transformed to a more robust set of parameters known as cepstral coefficients.

**Performance Analysis**
Following parameters are involved in performance evaluation of LPC's
- Bit Rates
- Overall Delay of the System
- Computational Complexity
- Objective Performance Evaluation

**Types of LPC**
Following are the types of LPC
- Voice-excitation LPC
- Residual Excitation LPC
- Pitch Excitation LPC
- Multiple Excitation LPC(MPLPC)
- Regular Pulse Excited LPC(RPLP)
- Coded Excited LPC(CELP)

### V. MEL FREQUENCY CEPSTRAL COEFFICIENTS (MFCC)

The use of Mel Frequency Cepstral Coefficients can be considered as one of the standard method for feature extraction (Motlíček, 2002). The use of about 20 MFCC coefficients is common in ASR, although 10-12 coefficients are often considered to be sufficient for coding speech (Hagen at al., 2003). The most notable downside of using MFCC is its sensitivity to noise due to its dependence on the spectral form. Methods that utilize information in the periodicity of speech signals could be used to overcome this problem, although speech also contains aperiodic content (Ishizuka & Nakatani, 2006).

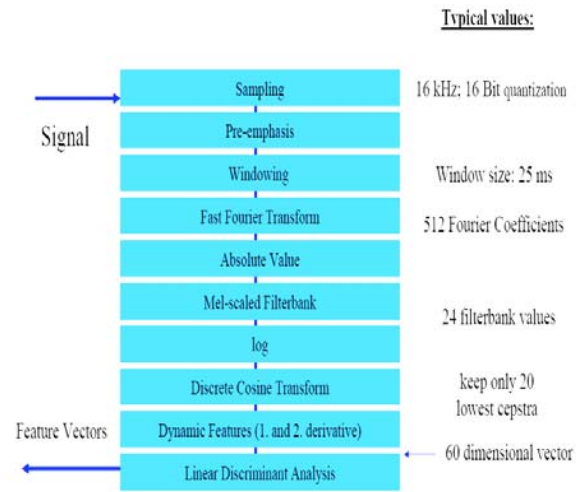

**Fig : MFCC : Complete Pipeline MFCC**

**Methodology**
The non-linear frequency scale used an approximation to the Mel-frequency scale which is approximately linear for frequencies below 1 kHz and logarithmic for frequencies above 1 kHz. This is motivated by the fact that the human auditory system becomes less frequency-selective as frequency increases above 1 kHz.

The MFCC features correspond to the cepstrum of the log filterbank energies. To calculate them, the log energy is first computed from the filterbank outputs as

$$S_t[m] = \ln\left(\sum_{n=0}^{N-1} |X_t[n]|^2 H_m[n]\right) \qquad 0 \leq m < M,$$

where $X_t[n]$ is the DFT of the $t^{th}$ input speech frame, $H_m[n]$ is the frequency response of $m^{th}$ filter in the filterbank, N is the window size of the transform and M is the total number of filters. Then, the discrete cosine transform (DCT) of the log energies is computed as

$$\vec{c}_t[m] = \sum_{n=0}^{M-1} S_t[n] \cos\left(\pi m \left(\frac{n-0.5}{M}\right)\right) \qquad 0 \leq m < M.$$

Since the human auditory system is sensitive to time evolution of the spectral content of the signal, an effort is often made to include the extraction of this information as part of feature analysis. In order to capture the changes in the coefficients over time, first and second difference coefficients are computed as respectively.

$$\Delta \vec{c}_t = \vec{c}_{t+2} - \vec{c}_{t-2}$$

$$\Delta\Delta \vec{c}_t = \Delta \vec{c}_{t+1} - \Delta \vec{c}_{t-1}$$

These dynamic coefficients are then concatenated with the static coefficients $\vec{c}_k$ according to making up the final output of feature analysis representing the t$^{th}$ speech frame.

$$\vec{x}_t = \begin{bmatrix} \vec{c}_t & \Delta \vec{c}_t & \Delta\Delta \vec{c}_t \end{bmatrix}^T$$

## VI. LFCC SPEECH FEATURES (LFCC-FB40)

The LFCC is computed as the MFCC-FB40 with the only difference that the Mel-frequency warping step is skipped. Thus, the desired frequency range is implemented by a filter-bank of 40 equal-width and equal-height linearly spaced filters. The bandwidth of each filter is 164 Hz, and the whole filter-bank covers the frequency range [133, 6857] Hz. Obviously, the equal bandwidth of all filters renders unnecessary the effort for normalization of the area under each filter.

**Computation of the LFCC**
- The N - point DFT is applied on the discrete time domain input signal x(n).
- The filter bank is applied on the magnitude spectrum [absolute value of x (k)]
- The logarithmically compressed filter-bank outputs [X.sub.i] are computed.
- Finally, the DCT is applied on the filter-bank outputs to obtain the LFCC FB-40 parameters.

Analogically to the MFCC FB-40 we compute only the first J = 13 parameters.

## VII. HFCC-E of Skowronsky & Harris:

[Skowronski & Harris,2002] introduced the Human Factor Cepstral Coefficients (HFCC-E). In the HFCC-E scheme the filter bandwidth is decoupled from the filter spacing. This is in contrast to the earlier MFCC implementations, where these were dependent variables. Another difference to the MFCC is that in HFCC-E the filter bandwidth is derived from the equivalent rectangular bandwidth (ERB), which is based on critical bands concept introduced by Moore and Glasberg, 1995 rather than on the Mel scale. Still, the centre frequency of the individual filters is computed by utilizing the Mel scale. Furthermore, in HFCC-E scheme the filter bandwidth is further scaled by a constant, which Skowronski and Harris labeled as E-factor. Larger values of the E-factor E={4, 5, 6} were reported.

## VIII. PURE FFT

Despite the popularity of MFCCs and LPC, direct use of vectors containing coefficients of FFT power-spectrum are also possible for feature extraction. As compared to methods exploiting knowledge about the human auditory system, the pure FFT spectrum carries comparatively more information about the speech signal. However, much of the extra information is located at the relatively higher frequency bands when using high sampling rates (e.g., 44.1 kHz etc.), which are not usually considered to be salient in speech recognition. The logarithm of the FFT spectrum is also often used to model loudness perception.

## IX. POWER SPECTRAL ANALYSIS (FFT)

One of the more common techniques of studying a speech signal is via the power spectrum. The power spectrum of a speech signal describes the frequency content of the signal over time.

The first step towards computing the power spectrum of the speech signal is to perform a Discrete Fourier Transform (DFT). A DFT

computes the frequency information of the equivalent time domain signal. Since a speech signal contains only real point values, we can use a real-point Fast Fourier Transform (FFT) for increased efficiency. The resulting output contains both the magnitude and phase information of the original time domain signal.

## X. PERCEPTUAL LINEAR PREDICTION (PLP)

The Perceptual Linear Prediction PLP model developed by Hermansky 1990. The goal of the original PLP model is to describe the psychophysics of human hearing more accurately in the feature extraction process.
PLP is similar to LPC analysis, is based on the short-term spectrum of speech. In contrast to pure linear predictive analysis of speech, perceptual linear prediction (PLP) modifies the short-term spectrum of the speech by several psychophysically based transformations.

## XI. PLP SPEECH FEATURES (PLP-FB19)

The PLP parameters rely on Bark-spaced filter-bank of 18 filters for covering the frequency range [0, 5000] Hz. Specifically, the PLP coefficients are computed as follows:
- The discrete time domain input signal x(n) is subject to the N - point DFT
- The critical-band power spectrum is computed through discrete convolution of the power spectrum with the piece-wise approximation of the critical-band curve, where B is the Bark warped frequency obtained through the Hertz-to-Bark conversion.
- Equal loudness pre-emphasis is applied on the down-sampled
- Intensity-loudness compression is performed.
- The result obtained so far an inverse DFT is performed to obtain the equivalent autocorrelation function.
- Finally, the PLP coefficients are computed after autoregressive modeling and conversion of the autoregressive coefficients to cepstral coefficients.

## XII. MEL SCALE CEPSTRAL ANALYSIS (MEL)

Mel scale cepstral analysis is very similar to perceptual linear predictive analysis of speech, where the short term spectrum is modified based on psychophysically based spectral transformations. In this method, however, the spectrum is warped according to the MEL Scale, whereas in PLP the spectrum is warped according to the Bark Scale. The main difference between Mel scale cepstral analysis and perceptual linear prediction is related to the output cepstral coefficients. The PLP model uses an all-pole model to smooth the modified power spectrum. The output cepstral coefficients are then computed based on this model. In contrast Mel scale cepstral analysis uses cepstral smoothing to smooth the modified power spectrum. This is done by direct transformation of the log power spectrum to the cepstral domain using an inverse Discrete Fourier Transform (DFT).

## XIII. RELATIVE SPECTRA FILTERING (RASTA)

To compensate for linear channel distortions the analysis library provides the ability to perform RASTA filtering. The RASTA filter can be used either in the log spectral or cepstral domains. In effect the RASTA filter band passes each feature coefficient. Linear channel distortions appear as an additive constant in both the log spectral and the cepstral domains. The high-pass portion of the equivalent band pass filter alleviates the effect of convolutional noise introduced in the channel. The low-pass filtering helps in smoothing frame to frame spectral changes.

## XIV. RASTA-PLP

Another popular speech feature representation is known as RASTA-PLP, an acronym for Relative Spectral Transform - Perceptual Linear Prediction. PLP was originally proposed by Hynek Hermansky as a way of warping spectra to minimize the differences between speakers while preserving the important speech information [Herm90]. RASTA is a separate technique that applies a band-pass filter to the energy in each frequency subband in order to smooth over short-term noise variations and to remove any constant offset resulting from static spectral coloration in the speech channel e.g. from a telephone line [HermM94].

## XV. COMBINED LPC & MFCC
[K.R. Aida–Zade, C. Ardil and S.S. Rustamov, 2006]

The determination algorithms MFCC and LPC coefficients expressing the basic speech features are developed by author. Combined use of cepstrals of MFCC and LPC in speech recognition system is suggested by author to improve the reliability of speech recognition system. The recognition system is divided into MFCC and LPC-based recognition subsystems. The training and recognition processes are realized in both subsystems separately, and recognition system gets the decision being the same results of each subsystems. Author claimed that, results in decrease of error rate during recognition.

**Steps of Combined use of cepstrals of MFCC and LPC :**
1. The speech signals is passed through a first-order FIR high pass filter
2. Voice activation detection (VAD). Locating the endpoints of an utterance in a speech signal with the help of some commonly used methods such as short-term energy estimate $E_s$, short-term power estimate $P_s$, short-term zero crossing rate $Z_s$ etc.
3. Then the mean and variance for the measures calculated for background noise, assuming that the first 5 blocks are background noise.
4. Framing.
5. Windowing.
6. Calculating of MFCC features.
7. Calculating of LPC features.
8. The speech recognition system consists of MFCC and LPC-based two subsystems. These subsystems are trained by neural networks with MFCC and LPC features, respectively.

**The recognition process stages:**
1. In MFCC and LPC based recognition subsystems recognition processes are realized in parallel.
2. The recognition results of MFCC and LPC based recognition subsystems are compared and the speech recognition system confirms the result, which confirmed by the both subsystems.

Since the MFCC and LPC methods are applied to the overlapping frames of speech signal, the dimension of feature vector depends on dimension of frames. At the same time, the number of frames depends on the length of speech signal, sampling frequency, frame step, frame length. Author uses sampling frequency is 16khs, the frame step is 160 samples, and the frame length is 400 samples. The other problem of speech recognition is the same speech has different time duration. Even when the same person repeats the same speech, it has the different time durations. Author suggested that, for partially removing the problem, time durations are led to the same scale. When the dimension of scale defined for the speech signal increases, then the dimension of feature vector corresponding to the signal also increases.

## XVI. MATCHING PURSUIT (MP)
[Selina Chu, Shrikanth Narayanan, and Jay Kuo, 2009]

Desirable types of features should be robust, stable, and straightforward, with the representation being sparse and physically interpretable. Author explained that using MP this representation is possible. The advantages of this representation are the ability to capture the inherent structure within each type of signal and to map from a large, complex signal onto a small, simple feature space. More importantly, it is conceivably more invariant to background noise and could capture characteristics in the signal where MFCCs tend to fail.

MP is a desirable method to provide a coarse representation and to reduce the residual energy with as few atoms as possible. Since MP selects atoms in order by eliminating the largest residual energy, it lends itself in providing the most useful atoms, even just after a few iterations.

The MP algorithm selects atoms in a stepwise manner among the set of waveforms in the dictionary that best correlate the signal structures. The iteration can be stopped when the coefficient associated with the atom selection falls below a threshold or when a certain number of atoms selected overall have been reached. Another common stopping criterion is to use the signal to residual energy ratio.

MP features are selected by the following process.
1. Windowing
2. Decomposition
3. Process stops after obtaining atoms.
4. Record the frequency and scale parameters for each of these atoms
5. Find the mean and the standard deviation corresponding to each parameter separately, resulting in 4 feature values.
6. Chose atoms to extract features for both training and test data.
7. The robustness of these features is enhanced by averaging.

MP selects vector that is exactly in the order of eliminating the largest residual energy. That is even the first few atoms found by MP will naturally contain the most information, making them to be more significant features. This also allows to map each signal from a larger problem space into a point in a smaller feature space.

## XVII. INTEGRATED PHONEME SUBSPACE (IPS)
[Hyunsin Park, Tetsuya Takiguchi, and Yasuo Ariki, 2009]

The effectiveness of feature extraction methods has been confirmed in speech recognition or speech enhancement experiments, but it remains difficult to recognize observed speech in reverberant environments. If the impulse response of a room is longer than the length of short-time Discrete Fourier Transform (DFT), the effects of reverberation are both additive and multiplicative in the power spectrum domain. Consequently, it becomes difficult to estimate the reverberant effects in the time or frequency domain. Therefore, author proposed a new data driven speech feature extraction method that is "Integrated Phoneme Subspace (IPS) method", which is based on the logarithmic mel-frequency filter bank domain.

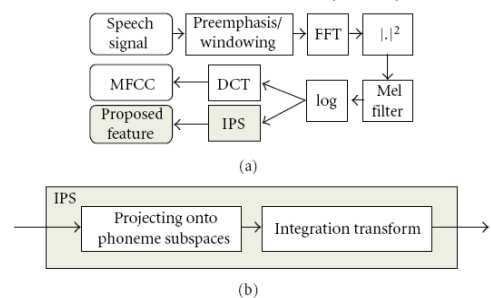

**Figure IPS: Block diagrams**
(a) Feature extraction of MFCC and proposed feature, (b) Integrated Phoneme Subspace (IPS) transform.

From the experimental work the proposed feature is obtained by transform matrices that are linear and time-invariant. The MDL-

based phoneme subspace selection experiment confirmed that optimal subspace dimensions differ. Author claimed that, the experiment results in isolated word recognition under clean and reverberant conditions showed that the proposed method outperforms conventional MFCC. The proposed method can be combined with other methods, such as speech signal processing or model adaptation, to improve the recognition accuracy in real-life environments.

## XVIII. CONCLUDING HIGHLIGHTS
- The feature space of a MFCC obtained using DCT is not directly dependent on speech data, the observed signal with noise does not show good performance without utilizing noise suppression methods.
- ICA, applied to speech data in the time or time-frequency domain, it gives good performance in phoneme recognition tasks.
- LDA, applied to speech data in the time-frequency domain shows better performance than combined linear discriminants in the temporal and spectral domain in continuous digit recognition task.
- MEL analysis and PLP analysis of speech, are similar where the short term spectrum is modified based on psychophysically based spectral transformations.
- In MEL the spectrum is warped according to the MEL Scale.
- In PLP the spectrum is warped according to the Bark Scale.
- The PLP model uses an all-pole model to smooth the modified power spectrum. The output cepstral coefficients are then computed.
- Mel scale cepstral analysis uses cepstral smoothing to smooth the modified power spectrum. This is done by direct transformation of the log power spectrum to the cepstral domain using an inverse DFT.
- In LPC reduced word error rates is found in difficult conditions as compared to PLP.
- FFT-based approach is good for its linearity in the frequency domain and its computational speed.
- FFT does not discard or distort information in any anticipatory manner.
- In FFT the representation of the signal remains easily perceivable for further analysis and post-processing.
- The effects of noise in the FFT spectrum can also be easily comprehended.
- The PLP features outperform MFCC in specific conditions.
- The MFCC reduces the frequency information of the speech signal into a small number of coefficients.
- In MFCC, the logarithmic operation attempts to model loudness perception in the human auditory system.
- MFCC is a very simplified model of auditory processing; it is easy and relatively fast to compute.
- The PLP functions provides limited capability of dealing with these distortion by employing a RASTA filter which makes PLP analysis more robust to linear spectral distortions

## XIX. CONCLUSION
We have discussed some features extraction techniques and their pros and cons. Some new methods are developed using combination of more techniques. Authors have claimed improvement in performance. There is a need to develop new hybrid methods that will give better performance in robust speech recognition area.